\documentclass[useAMS,usenatbib]{mn2e}

\include{epsf}

\def\gsim{\lower.73ex\hbox{$\sim$}\llap{\raise.4ex\hbox{$>$}}$\,$}
\def\lsim{\lower.73ex\hbox{$\sim$}\llap{\raise.4ex\hbox{$<$}}$\,$}
\def\mpc{$\,h^{-1}\,$Mpc}
\def\%{~per~cent}

\title[New H-band Galaxy Number Counts]
{New H-band Galaxy Number Counts: A Large Local Hole in the Galaxy Distribution?}
\author[W.J. Frith, N. Metcalfe \& T. Shanks]
{W.J. Frith\thanks{E-mail:w.j.frith@durham.ac.uk}, 
N. Metcalfe \& T. Shanks\\
Dept. of Physics, Univ. of Durham, South Road, Durham DH1 3LE, UK}
\begin{document}

\date{Accepted 2005. Received 2005; in original form 2005 }

\pagerange{\pageref{firstpage}--\pageref{lastpage}} \pubyear{2005}

\maketitle

\label{firstpage}

\begin{abstract}

We examine $H$-band number counts determined using new photometry over two fields with a combined solid angle of 0.30 deg$^2$ to $H\approx19$, as well 
as bright data ($H\le 14$) from the 2 Micron All 
Sky Survey (2MASS). First, we examine the bright number counts from 2MASS extracted for the $\approx$4000 deg$^2$ APM survey area situated around the 
southern galactic pole. We find a deficiency of $\approx$25\% at $H=13$ with respect to homogeneous predictions, in line with previous 
results in the $B$-band and $K_s$-band. In addition we examine the bright counts extracted for $|b|>$20$^{\circ}$ (covering $\approx$27$\,$000 
deg$^2$); we find a relatively constant deficit in the counts of $\approx$15-20\% to $H=14$. We investigate various possible causes for these 
results; namely, errors in the model normalisation, unexpected luminosity evolution (at low and high redshifts), errors in the photometry, 
incompleteness and large-scale structure. In order to address the issue of the model normalisation, we examine the number counts determined for the 
new faint photometry presented in this work and also for faint data ($H$\lsim 20) covering 0.39 deg$^2$ from the Las Campanas Infra Red Survey 
(LCIRS). In each case a zeropoint is chosen to match that of the 2MASS photometry at bright 
magnitudes using several hundred matched point sources in each case. We find a large offset between 2MASS and the LCIRS data of 0.28$\pm$0.01 
magnitudes. Applying a consistent zeropoint, the faint data, covering a combined solid angle of 0.69 deg$^2$, is in good agreement with the 
homogeneous prediction used previously, with a best fit normalisation a factor of 1.095$_{-0.034}^{+0.035}$ higher. We examine possible effects 
arising from unexpected galaxy evolution and photometric errors and find no evidence for a significant contribution from either. However, 
incompleteness in the 2MASS catalogue ($<10$\%) and in the faint data (likely to be at the few \%~level) may have a significant contribution. 
Addressing the contribution from large-scale structure, we estimate the cosmic variance in the bright counts over 
the APM survey area and for $|b|>$20$^{\circ}$ expected in a $\Lambda$CDM cosmology using 27 mock 2MASS catalogues constructed from the $\Lambda$CDM 
Hubble Volume simulation. Accounting for the model normalisation uncertainty and taking an upper limit for the effect arising from incompleteness, 
the APM survey area bright counts are in line with a rare fluctuation in the local galaxy 
distribution of $\approx2.5\sigma$. However, the $|b|>$20$^{\circ}$ counts represent a $4.0\sigma$ fluctuation, and imply a local hole which extends 
over the entire local galaxy distribution and is at odds with $\Lambda$CDM. The increase in faint near infrared data from the UK Infrared Deep Sky 
Survey (UKIDSS) should help to resolve this issue.

\end{abstract}


\section{Introduction}

A recurring problem arising from the study of bright galaxy number counts has been the measured deficiency of galaxies around the southern galactic 
pole. This was first examined in detail by \citet{sha} and subsequently by the APM galaxy survey \citep{mad}, which observed a large deficit in the 
number counts ($\approx$50\% at $B=$16, $\approx$30\% at $B=$17) over a $\approx$4000 deg$^2$ solid angle. If this anomaly was due solely to features 
in the galaxy distribution, this would be at odds with recent measurements of the variance of local galaxy density fluctuations 
\citep[e.g.][]{haw,fri2,cole3} or the expected linear growth of density inhomogeneities at large scales. 

\citet{mad3} examined possible causes of this deficiency. From redshift survey results over the APM survey area \citep{lov3}, it was argued that a 
weak local under-density contributed to the observed deficiency at the \lsim 10\% level at $B\approx$17. Instead, \citet{mad3} suggested that strong 
low redshift galaxy evolution was the dominant contribution. This phenomenon has also been suggested as a possible 
explanation for large deficiencies in the Sloan Digital Sky Survey \citep[SDSS;][]{lov2}, although models without such strong low redshift evolution 
provide predictions consistent with observed number redshift distributions \citep[e.g.][]{bro,col2,haw}. In contrast, \citet{sha} argued that 
evolution could not account for the observed slope and that large-scale structure was the principal cause of the deficiency in the counts.

However, another possible contribution to the low counts might be errors in the APM photometry. Comparing the photographic APM photometry with 
$B$-band CCD data, \citet{met} detected a small residual scale error in the APM survey zeropoints for $B$\gsim17. Correcting for this offset, the 
counts were now in good agreement with homogeneous predictions at faint magnitudes ($B$\gsim 17.5); however, the problematic deficiency at brighter 
magnitudes remained. More recently, \citet{bus} used $B$-band CCD data over $\approx$337 deg$^2$ within the APM survey area to provide the most 
accurate comparison to date with a sample of the APM survey photometry. The photometric zeropoint of this CCD data was in excellent agreement with 
the Millennium Galaxy Catalogue \citep{dri} and the Sloan Digital Sky Survey Early Data Release \citep{yas}. However, a comparison with the APM 
photometry suggested a large offset of 0.31 magnitudes for $B<$17.35. Applying this to the APM survey counts, a deficiency of $\approx$25\% remained 
at $B=$16; \citet{bus} determined that such a deficiency in the local galaxy distribution would still be at odds with a $\Lambda$CDM form to the 
galaxy correlation function and power spectrum at large scales.

In order to examine this issue independently, bright number counts have also been examined in the near infrared \citep{fri,fri3,fri4}. These 
wavelengths are particularly useful for such analysis as the number count predictions are fairly insensitive to the evolutionary model or the assumed 
cosmology at bright magnitudes (see Fig.~\ref{fig:fig1}); current observations are in remarkable agreement with predictions in the $K$-band to 
$K\approx$23 for example \citep{mcc} In particular, \citet{fri4} examined $K_s$-band number counts selected from the 2 Micron All Sky Survey 
\citep[2MASS;][]{jar2}. First, the counts over the APM survey area were determined; a similar deficiency was observed to the APM survey counts 
(with the zeropoint offset determined by \citet{bus} applied), with a $\approx$25\% deficit at $K_s=$12 compared to the no evolution model of 
\citet{met2}. Using a $\Lambda$CDM form for the angular correlation function at large scales and assuming the observed counts were solely due to 
features in the local galaxy distribution, the observed counts represented a $5\sigma$ fluctuation. However, this result was complicated by the fact 
that the 2MASS $K_s$-band number counts for almost the entire survey ($|b|>$20$^{\circ}$, covering $\approx$27$\,$000 deg$^2$) were also low, with a 
constant deficiency of $\approx$20\% between $K_s=$10 and $K_s=$13.5.

Did this surprising result perhaps indicate that the $K_s$-band \citet{met2} model normalisation was too high? Or, as suggested previously, could low 
redshift luminosity evolution significantly affect the bright counts? These issues were also addressed by \citet{fri4}: First, the \citet{met2} model 
was compared with faint $K$-band data collated from the literature. Fitting in the magnitude range 14$<K<$18 it was found that the best fit model 
normalisation was slightly too high, although not significantly (this magnitude range was used so as to avoid fluctuations in the counts arising from 
large-scale structure at bright magnitudes and significant effects from galaxy evolution at the faint end). Accounting for the normalisation 
uncertainty (of $\pm$6\%) the observed deficiency in the $K_s$-band counts over the APM survey area still represented a $\approx3\sigma$ fluctuation. 
Second, the issue of low redshift luminosity evolution was also addressed: 2MASS galaxies below $K_s=13.5$ were matched with the Northern and 
Southern areas of the 2dF Galaxy Redshift Survey \citep[2dGRS;][]{col}. The resulting $n(z)$, covering $>1000$ deg$^2$ in total, was consistent with 
the no evolution model of \citet{met2}. In addition, these $K_s$-band redshift distributions were used to form predictions for the number counts over 
the Northern and Southern 2dFGRS areas respectively. This was done by multiplying the luminosity function parameter $\phi^*$ (which governs the model 
normalisation) used in the \citet{met2} model by the relative density observed in the $K_s$-band $n(z)$ as a function of redshift. These  `variable 
$\phi^*$ models' were then compared with 2MASS counts extracted for the 2dFGRS areas in order to determine whether the observed counts were consistent   
with being due solely to features in the local galaxy distribution; the variable $\phi^*$ models were in good agreement with the number counts,  
indicating that low redshift luminosity evolution is unlikely to have a significant impact on the observed deficiency in the counts, in the 
$K_s$-band at least.

In this paper we aim to address the issue of low, bright number counts in the near infrared $H$-band. In particular we wish to address a drawback 
to the $K_s$-band analysis of \citet{fri4} - the issue of the number count model normalisation; while the $K_s$-band model used was compared with 
faint data and was found to be in good agreement, the level to which systematic effects, arising perhaps via zeropoint offsets between the bright and 
faint data or cosmic variance in the faint data, might affect the conclusions were uncertain. We address this issue in the $H$-band using new faint 
data covering 0.3 deg$^2$ to $H=18$, calibrated to match the 2MASS zeropoint. In section 2, we first verify that the $H$-band counts provide 
number counts over the APM survey area which are consistent with the previous results in the $B$ and $K_s$-bands \citep{bus,fri4}, and that the form 
of the counts is not significantly affected by low redshift luminosity evolution through comparisons with the variable $\phi^*$ models described 
above. In section 3, we provide details of the data reduction of the new faint $H$-band photometry. The associated counts are presented in section 4. 
In section 5 we discuss possible systematics affecting the bright number counts including the model normalisation and incompleteness. The conclusions 
follow in sections 6.

\section{Bright $H$-band counts from 2MASS}

\begin{figure}
\begin{center}
\centerline{\epsfxsize = 3.3in
\epsfbox{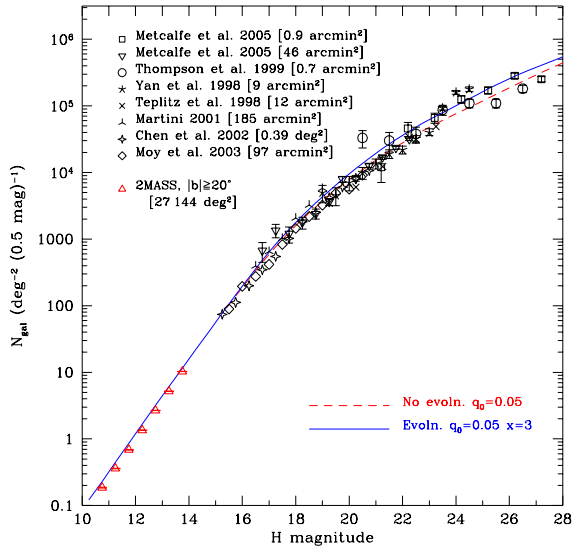}}
\caption{$H$-band galaxy number counts collated from the literature. The dashed and solid lines indicate the no evolution and pure luminosity
evolution predictions respectively, described in section 2. We also show bright $H$-band counts extracted from the 2MASS
extended source catalogue for $|b|>$20$^{\circ}$. For each dataset, we indicate the associated observed solid angle in square brackets.}
\label{fig:fig1}
\end{center}
\end{figure}

\begin{figure}
\begin{center}
\centerline{\epsfxsize = 3.3in
\epsfbox{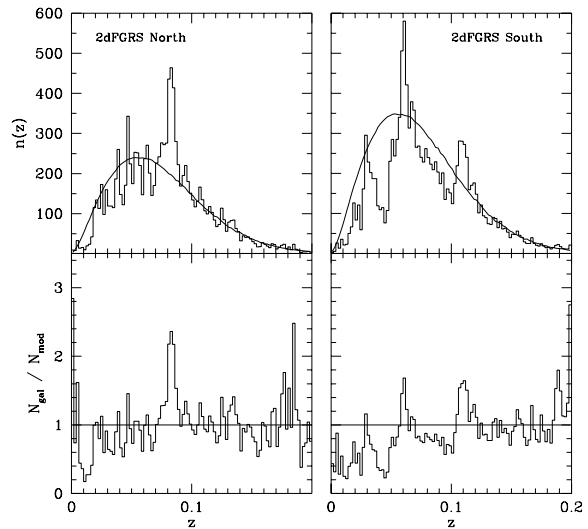}}
\caption{Number redshift histograms for 11$\,$501 and 13$\,$687 $H<$14 2MASS galaxies matched with the 446 deg$^2$ Northern (left hand) and
647 deg$^2$ Southern (right hand panels) 2dFGRS declination strips respectively. In each case the solid lines indicate the pure luminosity evolution
prediction for a homogeneous distribution described in section 2 normalised by the respective solid angles. We also indicate the relative density in 
the lower panels, dividing the observed $n(z)$ by the homogeneous prediction.}
\label{fig:fig2}
\end{center}
\end{figure}

We wish to examine the form of bright number counts in the $H$-band in order to verify that the counts over the APM 
survey area ($\approx$4000 deg$^2$ around the southern galactic pole) are comparable to those measured previously in the optical $B$-band and near 
infrared $K_s$-band \citep{bus,fri4}. The near infrared has the advantage of being sensitive to the underlying stellar mass and is much less 
affected by recent star formation history than optical wavelengths. For this reason, number count predictions in the near infrared are 
insensitive to the evolutionary model at bright magnitudes. In Fig.~\ref{fig:fig1} we show faint $H$-band data collated from the literature along 
with bright counts extracted from 2MASS over $\approx$27$\,$000 deg$^2$. The 2MASS magnitudes are determined via the 2MASS $H$-band extrapolated 
magnitude; this form of magnitude estimator has previously been shown to be an excellent estimate of the total flux in the $K_s$-band 
\citep{fri2,fri5} through comparison with the total magnitude estimates of \citet{jon} and the $K$-band photometry of \citet{lov}. Throughout this 
paper we use 2MASS $H$-band counts determined via this magnitude estimator. We also show two models in 
Fig.~\ref{fig:fig1} corresponding to homogeneous predictions assuming no evolution and pure luminosity evolution models. These are constructed from 
the $H$-band luminosity function parameters listed in \citet{met3} and the $K+E$-corrections of \citet{bru}. At bright magnitudes the two are 
indistinguishable; only at $H$\gsim 18 do the model predictions begin to separate. The faint data is in good agreement with both the no evolution and 
pure luminosity evolution predictions to $H\approx$26.

Before examining the $H$-band counts over the APM survey area, we first verify that the bright counts are consistent with relatively insignificant 
levels of low redshift luminosity evolution in the manner carried out by \citet{fri4} for the $K_s$-band counts. In the upper panels 
of Fig.~\ref{fig:fig2} we show $H$-band $n(z)$ to the 2MASS limiting magnitude of $H=14$, determined through matched 2MASS and 2dFGRS galaxies 
over the 2dFGRS Northern (left hand) and Southern (right hand panels) declination strips (see \citet{fri4} for further details of the 
matching technique). The solid lines indicate the expected homogeneous distribution constructed from the pure luminosity evolution predictions of 
\citet{met3} (there is no discernible difference between this and the no evolution prediction). In the lower panels we divide through by this 
prediction; these panels show the relative density as a function of redshift. The observed $n(z)$ are consistent with the expected trends, with 
relatively homogeneous distributions beyond $z=0.1$ (1\% and 8\% over-dense in the North and South respectively for $0.1\le z\le0.2$). For this reason, 
Fig.~\ref{fig:fig2} suggests that the level of luminosity evolution is relatively insignificant at low redshifts in the $H$-band; strong luminosity 
evolution produces an extended tail in the predicted $n(z)$ which is not observed in the data.

\begin{figure}
\begin{center}
\centerline{\epsfxsize = 3.3in
\epsfbox{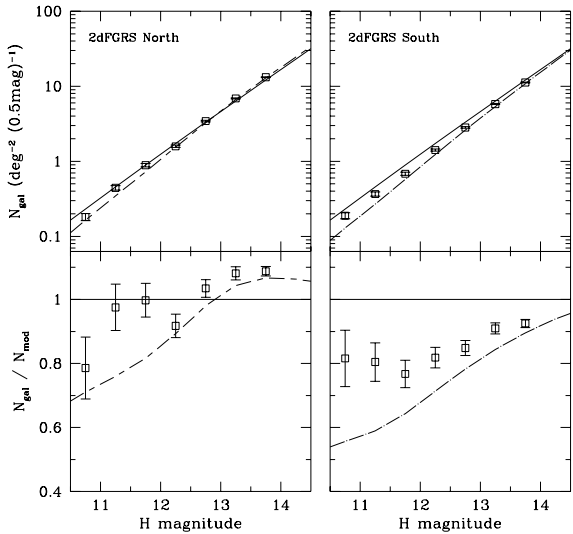}}
\caption{$H$-band 2MASS galaxy number counts extracted from the Northern (left hand) and Southern (right hand) 2dFGRS declination strips. The solid
line indicates the homogeneous pure luminosity evolution prediction described in section 2 (this and the no evolution prediction are 
indistinguishable at these magnitudes). The dashed and dot-dashed lines indicate the variable $\phi^*$ models for the Northern and Southern 2dFGRS 
strips respectively; these indicate the expected number counts given the observed $n(z)$ (Fig.~\ref{fig:fig2}). In the lower panels we divide through 
by the homogeneous prediction. In each case the errorbars indicate the Poisson uncertainty in each bin.}
\label{fig:fig3}
\end{center}
\end{figure}

\begin{figure}
\begin{center}
\centerline{\epsfxsize = 3.3in
\epsfbox{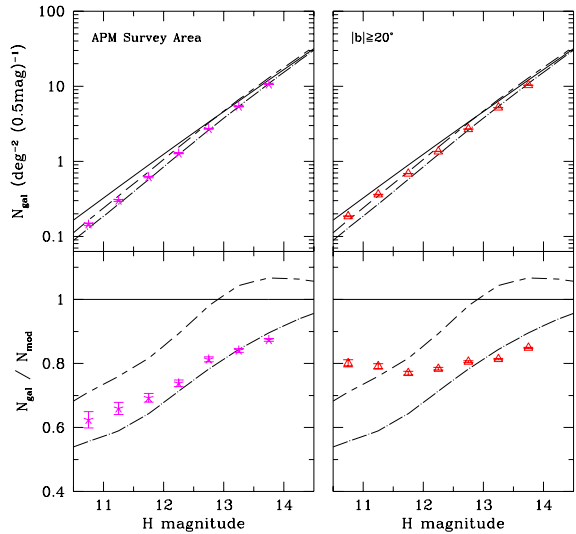}}
\caption{$H$-band 2MASS galaxy number counts extracted for the APM survey area ($\approx$4$\,$000 deg$^2$) and for $|b|>$20$^{\circ}$
($\approx$27$\,$000 deg$^2$), shown in the left and right hand panels respectively. As in Fig~\ref{fig:fig2}, we show the homogeneous pure 
luminosity prediction (solid line), and the Northern (dashed) and Southern (dot-dashed) variable $\phi^*$ models, indicating the expected number 
counts for the redshift distributions shown in Fig.~\ref{fig:fig2}. As before, in the lower panels we divide through by the homogeneous prediction. In 
each case the errorbars indicate the Poisson uncertainty in each bin.}
\label{fig:fig4}
\end{center}
\end{figure}

As a further check against strong low redshift luminosity evolution, we can use the observed $n(z)$ to predict the expected $H$-band number counts 
over the 2dFGRS declination strips. This technique is described in detail in \citet{fri,fri4}. To recap, we use the observed density 
(Fig.~\ref{fig:fig2}, lower panels), to vary the luminosity function normalisation ($\phi^*$) used in the \citet{met3} model as a function of 
redshift (for $z\le0.2$). We show these `variable $\phi^*$ models' along with the 2MASS $H$-band counts extracted for the 2dFGRS strips in 
Fig.~\ref{fig:fig3}. In each case, the upper panels indicate the number count on a logarithmic scale; in the lower panels we divide through by the 
homogeneous prediction. In both the Northern and Southern 2dFGRS areas, the counts are in good agreement with the expected trend, defined by the 
corresponding variable $\phi^*$ model. This indicates that real features in the local galaxy distribution are the dominant factor in the form of the 
observed $H$-band number counts, and that strong low redshift luminosity evolution is unlikely to have a significant role in any under-density 
observed in the APM survey area.

We are now in a position to examine the number counts over the APM survey area. In Fig.~\ref{fig:fig4} we show counts extracted for the 
$\approx$4000 deg$^2$ field along with the homogeneous and the Northern and Southern 2dFGRS variable $\phi^*$ models shown in 
Fig.~\ref{fig:fig2}. The form of the counts is in good agreement with the $B$ \citep{bus} and $K_s$-band \citep{fri4} bright number counts measured 
over the APM survey area, with a deficiency of $\approx$25\% below $H=13$. In addition, the form of the counts is similar to that of the counts 
extracted from the 2dFGRS Southern declination strip and the corresponding variable $\phi^*$ model (this is also observed in the $B$ and 
$K_s$-band); this perhaps indicates that the form of the local galaxy distribution in the $\approx$600 deg$^2$ 2dFGRS Southern declination 
strip is similar to that of the much larger APM survey area, with an under-density of $\approx$25\% to $z=$0.1. However, the 2MASS $H$-band counts 
over almost the entire survey ($|b|>$20$^{\circ}$, $\approx$27$\,$000 deg$^2$) are also deficient (as are the $K_s$-band counts), with 
a relatively constant deficit of $\approx$15-20\% to $H=14$ (Fig~\ref{fig:fig3}, right hand panels). 

The low $|b|>$20$^{\circ}$ counts raise the question as to whether systematic effects are significant, or whether these counts are due to real 
features in the local galaxy distribution, as suggested by the agreement between the variable $\phi^*$ models and corresponding counts in 
Fig.~\ref{fig:fig3}. If the latter is true, then the size of the local hole would not only be much larger than previously suggested but would 
also represent an even more significant departure from the form of clustering at large scales expected in a $\Lambda$CDM cosmology. In the following 
two sections we address a possible source of systematic error using new faint $H$-band photometry - the model normalisation. Other possible 
causes for the low counts are also discussed in section 5.

\section{New faint $H$-band data}

\subsection{Observations \& data reduction}

Our data were taken during a three night observing run in September 2004 at
the f/3.5 prime focus of the Calar Alto 3.5m telescope in the Sierra
de Los Filabres in Andalucia, southern Spain. The $\Omega$-2000 infra-red
camera contains a $2048\times2048$ pixel HAWAII-2 Rockwell detector array, with
18.5$\mu$m pixels, giving a scale of 0.45''/pixel at the prime focus.
All observations were taken with the $H$-band filter.
Poor weather meant that only just over one night's worth of data were usable,
and even then conditions were not photometric.

Our primary objective was to image the William Herschel Deep Field as deeply as
possible (the results of which are presented in a forthcoming paper), but time was available at the start of each night
to image several 'random' fields for 15 minutes each. These were composed of
individual 3 second exposures, stacked in batches of 10 before readout.
A dithering pattern on the sky with a shift up to $\pm$25'' around the nominal centre was adopted.

\begin{figure}
\begin{center}
\centerline{\epsfxsize = 3.3in
\epsfbox{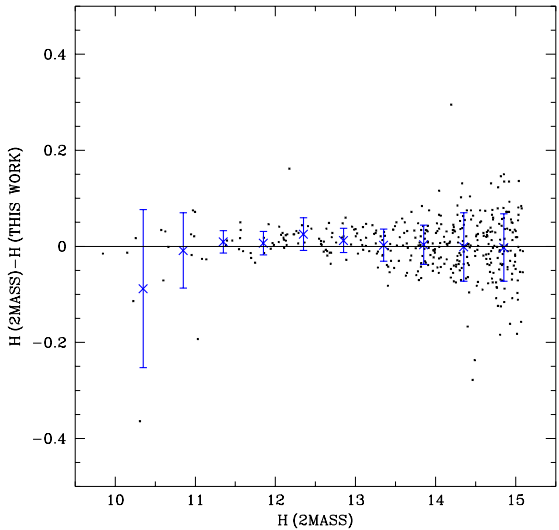}}
\caption{Here we examine the uncertainty in our photometric calibration with 2MASS. The $H$-band magnitudes determined by 2MASS and the
residual with our photometry are indicated for 393 point sources below $H=15.1$. The large datapoints indicate the mean offset and $rms$ dispersion as
a function of magnitude. The zeropoint used is indicated by the solid line and is accurate to $\pm$0.01 magnitudes at 1$\sigma$ confidence.}
\label{fig:fig5}
\end{center}
\end{figure}

\begin{figure}
\begin{center}
\centerline{\epsfxsize = 3.3in
\epsfbox{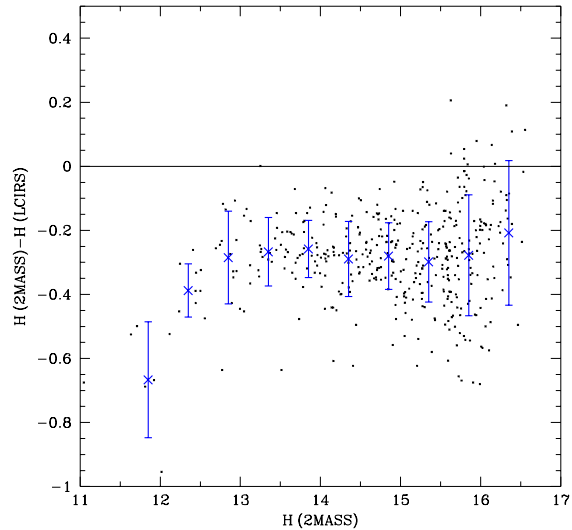}}
\caption{We compare the $H$-band photometry of the LCIRS \citep{che} with 2MASS using 438 points sources. As in Fig.~\ref{fig:fig5}, the large
datapoints indicate the mean offset and $rms$ dispersion as a function of magnitude. The mean offset is -0.28$\pm$0.01 magnitudes at 1$\sigma$
confidence. The zeropoint used in the new data presented in this work is indicated by
the solid line.}
\label{fig:fig6}
\end{center}
\end{figure}

Data reduction was complicated by the fact that both the dome and twilight
sky flat fields appeared to have a complicated out-of-focus pattern of the
optical train imprinted upon them (probably an image of the top end of the
telescope). This appeared (in reverse) in the science data if these frames
where used for flat-fielding. We therefore constructed a master flat field
by medianing together all the science frames from a particular night. This
was then used to flat-field all the data. Then, individual running medians were
constructed from batches of 10 or so temporally adjacent frames, and these
were subtracted from each frame to produce a flat, background subtracted
image. These were then aligned and stacked together (with sigma clipping to
remove hot pixels).

\subsection{Calibration}

Photometric calibration of the $H$-band images is obtained through comparison with the 2MASS point source catalogue. Fig.~\ref{fig:fig5} shows
the 2MASS magnitudes compared with our data for 393 matched point sources over the Calar Alto field and the William Herschel Deep Field. The 
zeropoint of our data is chosen to match that of the 2MASS objects and is accurate to $\pm$0.01 magnitudes. The large datapoints and errorbars 
indicate the mean offset and $rms$ dispersion as a function of magnitude. When comparing this data to the 2MASS number counts at bright magnitudes it 
is important to note that the 2MASS point source catalogue includes a maximum bias in the photometric zeropoint of $<$2\% around the sky (see the 
2MASS website).

\subsection{Star/Galaxy separation}

We use the Sextractor software to separate objects below $H=18$; for this magnitude limit, the associated STAR\_CLASS parameter provides a reliable
indicator of stars and galaxies. We identify 30.0\% as galaxies (CLASS\_STAR$<$0.1), 58.9\%  as stars (CLASS\_STAR$>$0.9), leaving 11.1\% as
unclassified.

\begin{table*}
\centering
\begin{tabular}{|c|c|c|c|c|c|c|}
\hline
$H$ & N$_{CA~field}$ & N$_{WHDF}$ & N$_{HDFS}$ & N$_{CDFS}$& N$_{tot}$ & N$_{mod}$ \\
  & & & & & (deg$^{-2}$ (0.5mag)$^{-1}$) & (deg$^{-2}$ (0.5mag)$^{-1}$) \\
\hline
14.25   & 10  & 4  & 6   & 8   & 40.8                   & 23.0 \\
14.75   & 17  & 5  & 12  & 8   & 61.1                   & 43.5 \\
15.25   & 21  & 9  & 23  & 23  & 110                    & 81.9 \\
15.75   & 41  & 14 & 31  & 43  & 188                    & 153 \\
16.25   & 55  & 15 & 77  & 51  & 288                    & 280 \\
16.75   & 133 & 39 & 163 & 73  & 594                    & 500 \\
17.25   & 217 & 58 & 238 & 135 & 943                    & 861 \\
17.75   & 283 & 77 & 337 & 256 & 1.44$\times$10$^3$     & 1.43$\times$10$^3$ \\
\hline
\end{tabular}
\caption{The raw number counts per half magnitude are shown for the new $H$-band data described in section 3 - the CA field (0.27 deg$^2$) and WHDF
(0.06 deg$^2$) in columns 2 and 3. In addition, we show the counts for the LCIRS fields, the HDFS (0.24 deg$^2$) and CDFS (0.16 deg$^2$) in columns 4
and 5, applying the zeropoint offset determined with respect to 2MASS in section 4.1. The total number count per deg$^2$ for all fields combined
(0.69 deg$^2$) is shown in column 6 along with the homogeneous pure luminosity evolution prediction of \citet{met3} in column 7. The faintest
magnitude bin for the CA field is slightly smaller (0.21 deg$^2$) than at brighter magnitudes; the combined solid angle for the faintest bin in column
6 is therefore 0.66 deg$^2$.}
\label{table:counts}
\end{table*}

\section{Faint $H$-band counts}

\subsection{Comparison with the LCIRS}

Before determining number counts for the new $H$-band data described in section 3, we first examine the photometry of the Las Campanas Infra-Red 
Survey \citep[LCIRS;][]{che}. The published data covers 847 arcmin$^2$ in the Hubble Deep Field South (HDFS) and 561 arcmin$^2$ in the Chandra Deep 
Field South (CDFS); the combined solid angle (0.39 deg$^2$) represents the largest $H$-band dataset for 14\lsim $H$\lsim 20. The associated number 
counts are $\approx$15\% below the homogeneous \citet{met3} predictions at $H=18$ (see Fig.~\ref{fig:fig1}). This is significant, as if the model 
normalisation was altered to fit, the deficiency in the 2MASS counts at bright magnitudes (Fig.~\ref{fig:fig3}) would become much less severe. 
However, various other surveys show higher counts, although over much smaller solid angles. With the LCIRS data in particular therefore, it is vital 
to ensure that the photometric zeropoint is consistent with the 2MASS data at bright magnitudes.

In Fig.~\ref{fig:fig6} we compare the LCIRS and 2MASS $H$-band photometry for 438 points sources matched over the HDFS and CDFS fields. There appears 
to be a large offset which is approximately constant for $K>$12. Using point sources matched at all magnitudes, we determine a mean offset of 
-0.28$\pm$0.01 magnitudes; this is robust to changes in the magnitude range and is consistent over both the HDFS and CDFS fields.

\subsection{New $H$-band counts}

\begin{figure}
\begin{center}
\centerline{\epsfxsize = 3.3in
\epsfbox{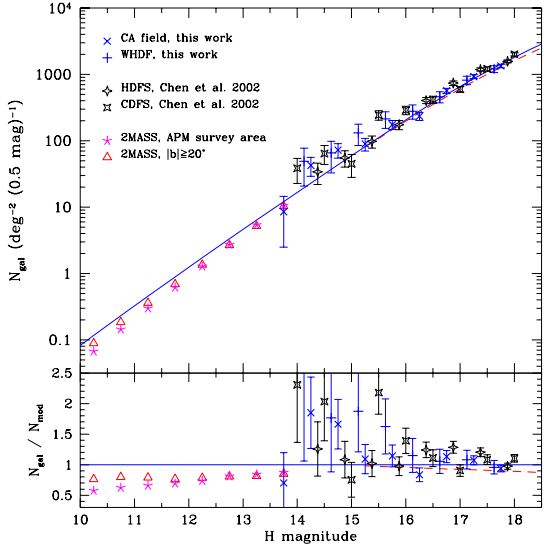}}
\caption{Here we show $H$-band galaxy number counts for the two separate fields observed in this work, the Calar Alto field (CA field; 0.27
deg$^2$) and the William Herschel Deep field (WHDF; 0.06 deg$^2$). We also show number counts determined for the two separate fields
of the LCIRS \citep{che} situated in the Hubble Deep Field South (HDFS; 0.24 deg$^2$) and Chandra Deep Field South (CDFS; 0.16 deg$^2$), subtracting
0.28 magnitudes in each case in order to bring the LCIRS and 2MASS zeropoints (and hence also the CA field and WHDF zeropoints) into
agreement. We also show bright number counts extracted from 2MASS for the APM survey area and for $|b|>$20$^{\circ}$ as shown in Fig.~\ref{fig:fig3}.
The models are indicated as in Fig.~\ref{fig:fig1}. In the lower panel, we divide through by the pure luminosity evolution homogeneous prediction as in
Figs.~\ref{fig:fig3} and \ref{fig:fig4}. At faint magnitudes, we indicate the Poisson uncertainty in each bin. We omit Poisson errors on the bright
counts for clarity (see Fig.~\ref{fig:fig4} for these). We discuss the uncertainty in the counts arising from cosmic variance in section 5.}
\label{fig:fig7}
\end{center}
\end{figure}

In Fig.~\ref{fig:fig7} we show counts determined for the new $H$-band data described in section 3, the 0.27 deg$^2$ CA field and the 0.06 deg$^2$ 
WHDF (see also table~\ref{table:counts}). Both sets of counts are in excellent agreement with the pure luminosity evolution and no evolution 
homogeneous predictions of \citet{met3}. In addition we show LCIRS counts determined in the 0.24 deg$^2$ HDFS and 0.16 deg$^2$ CDFS, applying the 
0.28 magnitude zeropoint offset determined with respect to 2MASS in section 4.1. The associated counts are also in excellent agreement with the 
\citet{met3} models at all magnitudes. 

\begin{figure}
\begin{center}
\centerline{\epsfxsize = 3.3in
\epsfbox{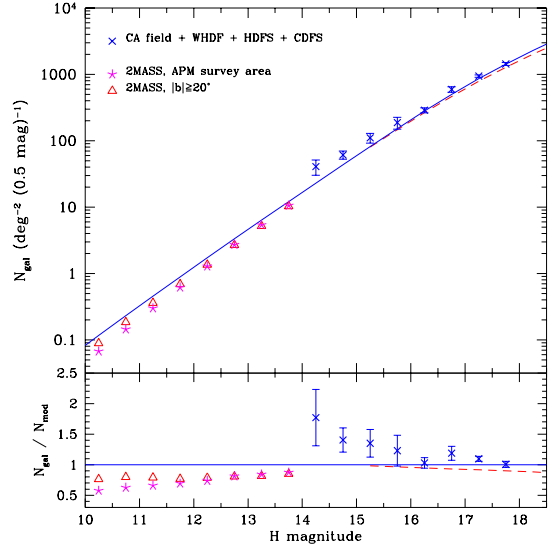}}
\caption{Here we show the faint $H$-band data from the two fields presented in this work (CA field and WHDF) and the two fields published by the LCIRS
\citep[HDFS and CDFS;][]{che}, applying a zeropoint to the LCIRS data consistent with the bright $H$-band 2MASS data (and hence the CA
field and WHDF also), as shown in Fig.~\ref{fig:fig7}. The errorbars at faint magnitudes indicate the field-to-field error, weighted in order to
account for the different solid angles of each field. Bright $H$-band counts extracted from 2MASS for the APM survey area and for $|b|>$20$^{\circ}$
are shown as previously. In the lower panel, the counts are divided through by the pure luminosity evolution homogeneous prediction as before. }
\label{fig:fig8}
\end{center}
\end{figure}

In Fig.~\ref{fig:fig8}, we show counts determined from our data and the LCIRS combined, with a consistent zeropoint applied as in Fig.~\ref{fig:fig7}.
We estimate the uncertainty arising from cosmic variance using field-to-field errors, weighted by the solid angle of each field. These combined counts 
are in good agreement with the \citet{met3} models, particularly at fainter magnitudes where the dispersion in the counts arising from cosmic 
variance appears to be small. We perform least squares fits between these counts and the pure luminosity evolution model; in the magnitude range 
14$<H<$18 we find a best fit normalisation of 1.095$_{-0.034}^{+0.035}$, where 1.0 corresponds to the \citet{met3} normalisation shown in 
Fig.~\ref{fig:fig8}. Varying the fitting range does slightly alter the result; in the range 16$<H<$18 we find a best fit normalisation of 
1.061$_{-0.033}^{+0.048}$ for example.

\section{Discussion}

In the previous sections, bright $H$-band number counts from 2MASS were determined over the APM survey area ($\approx4000$ deg$^2$) and almost the 
entire 66\% of the sky ($|b|>$20$^{\circ}$, $\approx$27$\,$000 deg$^2$), along with faint counts to $H=18$ over a combined solid angle of 0.69 
deg$^2$ applying a zeropoint consistent with 2MASS. The bright $H$-band number counts over the APM survey area are extremely low ($\approx25$\% at 
$H=13$) with respect to homogeneous predictions, and reproduce the form of the bright counts observed in the optical $B$-band \citep{bus} and the 
near infrared $K_s$-band \citep{fri4}. Previous work has suggested that, if due solely to local large-scale structure, these low counts would be at 
odds with the form of clustering expected in a $\Lambda$CDM cosmology. In addition, the bright $H$-band $|b|>$20$^{\circ}$ counts were also found to 
be low. In the following section, various possible causes for these low counts are examined.

\subsection{Model normalisation}

The normalisation of number count models may be determined by fixing the predicted to the observed number of galaxies at faint magnitudes. 
The magnitude range at which this is done should be bright enough to avoid large uncertainties in the evolutionary model while faint enough such 
that large fluctuations in the counts arising from cosmic variance are expected to be small. Near infrared wavelengths are expected to be insensitive 
to luminosity evolution at bright magnitudes, making the $H$-band particularly useful for such analysis. Of vital importance when determining the 
model normalisation is that when making comparisons between faint and bright counts, the zeropoints are consistent; an offset of a few tenths of a 
magnitude between the two, for example, would be enough to remove the observed anomaly in the bright counts over the APM survey area.

Applying the 2MASS zeropoint to the faint $H$-band data presented in this work and the LCIRS data \citep{che}, covering a combined solid angle of 
0.69 deg$^2$, it is clear that a discrepancy between the bright and faint counts exists; the model normalisation used previously, which indicates low 
counts below $H=14$ over the APM survey area (and for $|b|>$20$^{\circ}$), provides good agreement with the faint data. 
In fact, fixing the model to the faint counts implies a slightly higher normalisation. This agreement, as indicated by the errorbars in 
Fig.~\ref{fig:fig8}, suggests that the discrepancy between the bright and faint counts is not due to cosmic variance in the faint data. To remove the 
observed deficit in the APM survey area counts below $H=14$ by renormalising the model, requires a deviation from the faint counts of 7.0$\sigma$ 
using the best fit normalisation of 1.095$_{-0.034}^{+0.035}$ (determined for $14<H<18$). Similarly, renormalising to the $|b|>$20$^{\circ}$ 
counts would require a deviation of 7.2$\sigma$ from the faint data.

In addition, the model normalisation may also be scrutinised through comparison with redshift distributions. Fig.~\ref{fig:fig2} shows the 
\citet{met3} pure luminosity evolution model compared with $H$-band $n(z)$ determined through a match between 2MASS and the 2dFGRS Northern and 
Southern declination strips. The model predictions appear to be consistent with the observations, with relatively homogeneous distributions beyond 
$z=0.1$ (1\% and 8\% over-dense in the North and South respectively). Lowering the model normalisation to fit the bright 2MASS number counts would 
compromise this agreement and imply large over-densities beyond $z=0.1$ (19\% and 27\% in the North and South respectively).

\subsection{Galaxy evolution}

A change in amplitude therefore, cannot easily account for the discrepancy in the number counts at bright magnitudes. However, could an unexpected  
change in the slope of the number count model contribute? In section 2, we examined the consistency of the number counts at bright 
magnitudes with the underlying redshift distribution, assuming a model with insignificant levels of luminosity evolution at low redshift. The 
predictions derived from the observed $n(z)$ were in good agreement with the observed number counts indicating that luminosity evolution at low 
redshift is unlikely to have a significant impact on the form of the counts at bright magnitudes. This is supported by the consistency of the pure 
luminosity evolution model with the observed redshift distributions (Fig.~\ref{fig:fig2}); strong low redshift luminosity evolution produces a tail in 
the $n(z)$ which would imply large deficiencies at high redshift.

Could unexpectedly high levels of luminosity evolution at higher redshifts affect our interpretation of the bright counts? If the slope of the 
homogeneous prediction were to increase significantly above $H\approx14$ from the evolutionary models considered in this paper, then the model 
normalisation could effectively be lowered into agreement with the bright counts. The problem with this is that the number counts beyond $H\approx14$ 
are consistent with low levels of luminosity evolution to extremely faint magnitudes ($H\approx26$). Models with significantly higher levels of 
luminosity evolution above $H\approx14$ would therefore compromise this agreement.

Therefore, it appears that relatively low levels of luminosity evolution are consistent with number count observations to high redshifts. Also, recent 
evidence from the COMBO-17 survey, examining the evolution of early-type galaxies using nearly 5000 objects to $z\approx1$ \citep{bel2}, suggests that 
density evolution will also not contribute; $\phi^*$ appears to decrease with redshift indicating that the number of objects on the red sequence 
increases with time, and so acts contrary to the low counts observed at bright magnitudes. This picture is supported by the K20 survey \citep{cim}, 
which includes redshifts for 480 galaxies to a mean depth of ${\bar z}\approx0.7$ and a magnitude limit of $K_s=20$ with high completeness. The 
resulting redshift distribution is consistent with low levels of luminosity and density evolution \citep{met3}. 

In summary, significant levels of evolution are not expected in passive or star forming pure luminosity evolution models, although could occur through 
dynamical evolution. However, the pure luinosity evolution models of \citet{met3} fit the observed $H<14$ $n(z)$ at $z>0.1$; it is at lower redshifts 
that there are fluctuations. In addition, these models continue to fit the observed $n(z)$ at very high redshift and the number counts to extremely 
faint magnitudes ($K\approx23$), suggesting that there is little need for evolution at $z\approx1$, far less $z$\lsim 0.1. Some combination of 
dynamical and luminosity evolution might be able to account for these observations; however it would require fine-tuning in order to fit both the 
steep counts at bright magnitudes and the unevolved $n(z)$ at low and high redshifts.

\subsection{Photometry issues \& completeness}

The number counts shown in Figs.~\ref{fig:fig7} and \ref{fig:fig8} show bright and faint counts with a consistent zeropoint applied. Photometry 
comparisons have been made using several hundred point sources matched at bright magnitudes. In order to check that the applied zeropoints are 
consistent with the {\it galaxy} samples, we also compare the 2MASS photometry with 24 matched galaxies in the CA field and WHDF and 16 in the LCIRS 
samples; we find that the mean offsets are $-0.01\pm$0.04 and $-0.32\pm$0.06, consistent with the zeropoints determined via the 2MASS point sources. 
The comparisons with the 2MASS point source catalogue (Figs.~\ref{fig:fig5} and \ref{fig:fig6}) also indicate that there is no evidence of scale 
error in either of the faint samples to $H\approx16$.

Could the discrepancy between the bright and faint counts arise from an under-estimation of the total flux of the galaxies? Recall that we make no 
correction to total magnitude for the faint data presented in this work; however, under-estimating the total flux in the faint data would only 
increase the observed deficit in the counts at bright magnitudes, if the model normalisation is adjusted to fit the faint counts. The good agreement 
between the point source and galaxy zeropoints suggests that the estimate for the total galaxy flux is comparable in the bright and faint data. At 
bright magnitudes, the 2MASS extrapolated $H$-band magnitudes are used. In the $K_s$-band, this magnitude estimator has been shown to be an excellent 
estimate of the total flux, through comparisons with the total $K_s$-band magnitude estimator of \citet{jon} and the $K$-band photometry of 
\citet{lov}. 

Another possible contribution to the low counts could be high levels of incompleteness in the 2MASS survey. As with the possible systematic effects 
described previously, it is differing levels of completeness in the faint and bright data which would be important. The 2MASS literature quotes the 
extended source catalogue completeness as $>90$\% (see the 2MASS website for example). Independently, \citet{bel} suggest that the level of 
completeness is high ($\approx$99\%), determined via comparisons with the SDSS Early Data Release spectroscopic data and the 2dFGRS. The faint data 
presented in this work and the LCIRS data are likely to suffer less from incompleteness, as we cut well below the magnitude limit, are subject to 
lower levels of stellar confusion and suffer less from low resolution effects. Incompleteness in 2MASS will therefore affect the observed deficit 
in the bright counts at the $<10$\% level, although the effect is likely to be at the low end of this constraint due to incompleteness in the faint 
catalogues and suggestions that the 2MASS extended source catalogue is fairly complete.

\subsection{Large-scale structure}

It appears therefore, that the observed deficiency in the bright counts might be significantly affected by incompleteness in the 2MASS extended source 
catalogue. However, the level to which other systematic effects such as the model normalisation, luminosity evolution and photometry issues appears to 
be small. The question then is $-$ accounting for these various sources of error or uncertainty, are the deficiencies in the bright $H$-band counts 
over the APM survey area and for $|b|>20^{\circ}$ still at odds with the expected fluctuations in the counts arising from local large-scale structure 
in a $\Lambda$CDM cosmology, as suggested in previous work \citep{bus,fri4}?

We determine the expected fluctuations in the bright number counts due to cosmic variance via $\Lambda$CDM mock 2MASS catalogues; these 
are described in detail in \citet{fri4}. To recap, we apply the 2MASS selection function to 27 virtually independent volumes of
$r=500$\mpc~formed from the 3000$^3h^{-3}$Mpc$^3$ $\Lambda$CDM Hubble Volume simulation. This simulation has input parameters of
$\Omega_m=0.3$, $\Omega_b=0.04$, $h=0.7$ and $\sigma_8=0.9$ \citep{jen}. The mean number density of the counts at the magnitude limit is set to that
of the observed 2MASS density.

We are now in a position to estimate the significance of the observed bright $H$-band counts. We use the 1$\sigma$ fluctuation in the counts expected 
in a $\Lambda$CDM cosmology (determined using the 2MASS mocks described above), which for the APM survey area is 7.63\% (for $H<13$) and 4.79\% (for 
$H<14$), and for $|b|>20^{\circ}$ is 3.25\% (for $H<13$) and 1.90\% (for $H<14$). In addition we also take into account the uncertainty in the model 
normalisation; we use the best fit normalisation of the \citet{met3} pure luminosity evolution model (a factor of 1.095 above the
\citet{met3} model) and add the uncertainty of $\pm$3.1\% derived from the faint $H$-band counts (presented in Fig.~\ref{fig:fig8}) in quadrature.
Regarding the possible effect arising from survey incompleteness, we first assume that the level of incompleteness is comparable in the faint and 
bright data; the resulting significance for the APM survey area and $|b|>20^{\circ}$ bright counts are shown in column 3 of table~\ref{table:defsig}. 
This represents an upper limit on the significance since we have effectively assumed that there is no difference in the incompleteness between the 
bright and faint datasets. In column 4 of table~\ref{table:defsig}, we assume that there is a difference in the completeness 
levels in the faint and bright data of 10\%. This represents a lower limit on the significance (assuming that there are no further significant 
systematic effects), since we assume that the completeness of the 2MASS extended source catalogue is 90\% (the lower limit) and that there is no 
incompleteness in the faint data. 

Therefore, assuming a $\Lambda$CDM cosmology, it appears that the observed counts over the APM survey area might be in line with a rare 
fluctuation in the local galaxy distribution. However, the counts over 66\% of the sky ($|b|>20^{\circ}$) suggest a deficiency in the counts that are 
at odds with $\Lambda$CDM, even accounting for a 10\% incompleteness effect and the measured uncertainty in the best fit model normalisation.

\begin{table}
\centering
\begin{tabular}{|c|c|c|c|} 
\hline
Field 		& $H_{lim}$	& Significance  	& Significance  \\
      		&        	& (no incompleteness	& (assuming 10\% 	\\
		&		& correction)		& incompleteness) \\
\hline
APM   		& $13.0$  	& 3.7$\sigma$          	& 2.5$\sigma$   \\
APM   		& $14.0$  	& 4.2$\sigma$          	& 2.4$\sigma$   \\
\\
$|b|>20^{\circ}$& $13.0$ 	& 6.1$\sigma$       	& 3.8$\sigma$   \\
$|b|>20^{\circ}$& $14.0$ 	& 6.8$\sigma$       	& 4.0$\sigma$   \\
\hline
\end{tabular}
\caption{Here we show the significance of the $H$-band 2MASS counts extracted for the $\approx$4000 deg$^2$ APM survey area and for $|b|>20^{\circ}$, 
for $H<13$ and $H<14$. In each case we determine the expected cosmic variance using a $\Lambda$CDM form to the large-scale power determined via mocks 
constructed from the Hubble Volume simulation. In addition we use the best fit normalisation of the \citet{met3} pure luminosity evolution model 
determined at faint magnitudes (a factor of 1.095 above the original normalisation) and add the uncertainty on this ($\pm$3.1\%) in quadrature to the 
expected cosmic variance. In the third column we use the observed counts as shown in Figs.~\ref{fig:fig4}, \ref{fig:fig7} and \ref{fig:fig8}; in the 
fourth column we account for an upper limit on the incompleteness in the 2MASS extended source catalogue of 10\%; the level to which this will affect 
the significance is likely to be lower due to incompleteness in the faint data. } 
\label{table:defsig}
\end{table}

\section{Conclusions}

We have presented new $H$-band photometry over two fields with a combined solid angle of 0.30 deg$^2$ to $H\approx$19. The zeropoint is chosen to 
match that of the 2MASS photometry at the bright end and is accurate to $\pm$0.01 magnitudes. In addition we have examined the faint $H$-band data of 
the LCIRS \citep{che} which covers two fields with a combined solid angle of 0.39 deg$^2$ to $H\approx$20. The zeropoint of this data appears to be 
offset from the 2MASS photometry by 0.28$\pm$0.01 magnitudes. Applying a consistent zeropoint, the faint counts determined from the new data 
presented in this work and the LCIRS are in good agreement with the pure luminosity evolution model of \citet{met3}, with a best fit 
normalisation a factor of 1.095$_{-0.034}^{+0.035}$ higher.

In contrast, the bright $H$-band counts extracted from 2MASS over the $\approx$4000 deg$^2$ APM survey area around the southern galactic pole are low 
with respect to this model, corroborating previous results over this area in the optical $B$-band and near infrared $K_s$-band \citep{bus,fri4}. In 
addition, the counts extracted for almost the entire survey, covering 66\% of ths sky, are also low with a deficit of $15-20$\% to $H=14$. 
Importantly, this discrepancy does not appear to be due to zeropoint differences between the faint and bright data or uncertainty in the model 
normalisation set by the faint counts.

We have investigated various possible sources of systematic error which might affect this result: The counts are consistent with low levels of 
luminosity and density evolution, as predicted by the pure luminosity evolution model of \citet{met3}, to extremely faint magnitudes (see 
Fig.~\ref{fig:fig1}). Also, the photometry appears to be consistent between the faint and bright $galaxy$ data using a zeropoint applied via 
comparisons between point sources. However, differing incompleteness in the bright and faint galaxy samples might have a significant impact; 
completeness in the 2MASS extended source catalogue is $>90$\%. 

Finally, we determined the expected cosmic variance in the bright number counts from $\Lambda$CDM mock 2MASS catalogues. Allowing for the model 
normalisation uncertainty determined from the faint counts, and using an upper limit on the incompleteness in the 2MASS galaxy 
sample, the deficiency in the counts over the APM survey area represent a rare ($\approx$1 in 100) fluctuation in a $\Lambda$CDM cosmology. However, 
the low $H$-band counts for $|b|>20^{\circ}$ suggest that this deficiency might extend over the entire local galaxy distribution; allowing for 
incompleteness and the model normalisation uncertainty as before, this would represent a 4$\sigma$ fluctuation ($<$1 in 10$\,$000) in the local galaxy 
distribution, and would therefore be at odds with the expected form of clustering expected in a $\Lambda$CDM cosmology on large scales. The increase 
in faint near infrared data from the UK Infrared Deep Sky Survey (UKIDSS) should help to resolve this issue.

\section*{Acknowledgements} 

The new data presented in this paper were based on observations collected at the Centro Astron\`{o}mico Hispano Alem\'{a}n (CAHA) at Calar Alto, 
operated jointly by the Max-Planck Institut f\"{u}r Astronomie and the Instituto de Astrof\'{i}ca de Andaluc\'{i}a (CISC). This publication also 
makes use of data products from the 2 Micron All-Sky Survey, which is a joint project of the University of Massachusetts and the Infrared Processing 
and Analysis Centre/California Institute of Technology, funded by the Aeronautics and Space Administration and the National Science Foundation. We 
thank Peter Draper for his help with Sextractor. We also thank the wonderful Phil Outram and John Lucey for useful discussion and Nicholas 
Ross for assistance with the Calar Alto field selection.

\label{lastpage}

\end{document}